\documentclass[sigconf,natbib=true]{acmart}

\setcopyright{none}
\renewcommand\footnotetextcopyrightpermission[1]{}
\acmConference[AgentSearch@SIGIR'26]{AgentSearch Workshop at SIGIR 2026}{July 24, 2026}{Melbourne, Australia}

\usepackage{amsmath}
\usepackage{booktabs}
\usepackage{placeins}
\usepackage{xurl}
\usepackage{subcaption}
\usepackage{enumitem}

\newcommand{\MS}{Minority Sentinel}
\newcommand{\feat}[1]{\mbox{\texttt{#1}}}

\begin{document}

\title{Minority Sentinel: When to Overturn Majority Voting in Multi-Agent LLM Debates}

\author{Chuan He}
\email{chuan.he3@unsw.edu.au}
\affiliation{%
  \institution{University of New South Wales}
  \city{Sydney}
  \country{Australia}
}

\author{Zebin Chen}
\email{zebin.chen1@unsw.edu.au}
\affiliation{%
  \institution{University of New South Wales}
  \city{Sydney}
  \country{Australia}
}

\author{Zhengyi Yang}
\email{zhengyi.yang@unsw.edu.au}
\affiliation{%
  \institution{University of New South Wales}
  \city{Sydney}
  \country{Australia}
}

\author{Shaobo Qiao}
\email{shaobo.qiao@eulerai.au}
\affiliation{%
  \institution{Euler AI}
  \city{Sydney}
  \country{Australia}
}

\author{Mingchen Ju}
\email{mingchen.ju@eulerai.au}
\affiliation{%
  \institution{Euler AI}
  \city{Sydney}
  \country{Australia}
}

\author{Jiate Liu}
\email{jiate.liu@unsw.edu.au}
\affiliation{%
  \institution{University of New South Wales}
  \city{Sydney}
  \country{Australia}
}

\author{Dong Wen}
\email{dong.wen@unsw.edu.au}
\affiliation{%
  \institution{University of New South Wales}
  \city{Sydney}
  \country{Australia}
}

\author{Guanfeng Liu}
\email{guanfeng.liu@mq.edu.au}
\affiliation{%
  \institution{Macquarie University}
  \city{Sydney}
  \country{Australia}
}

\renewcommand{\shortauthors}{He et al.}

\begin{abstract}
Multi-Agent Debate (MAD) with Majority Voting is a dominant paradigm for improving LLM reasoning, yet its effectiveness rests on the Condorcet Jury Theorem's assumption of independent errors. Because contemporary LLMs share similar pretraining corpora, their errors are strongly correlated, causing the majority to systematically suppress correct minority opinions, a phenomenon we term \emph{Minority Truth}. Through debates among three heterogeneous LLM agents on six benchmarks, we find that roughly one in four divergent cases has the minority holding the correct answer, yielding a 10-percentage-point theoretical recovery margin. We propose \textbf{\MS{}}, a lightweight meta-classifier that extracts a multi-dimensional \emph{debate fingerprint} from debate logs and trains a LightGBM model to decide when to overturn majority voting. \MS{} achieves a stable Flip Precision of 81.2\% with positive Net Gain across all six datasets and all 20 random seed trials, demonstrating that debate logs contain sufficient behavioral signals for a non-LLM classifier to reliably recover suppressed minorities without degrading system accuracy. The LLM-as-Judge baseline yields negative Net Gain despite higher recall, confirming that flip safety, not recovery volume, determines intervention value.
\end{abstract}



\maketitle

\section{Introduction}

Multi-Agent Debate (MAD) has emerged as an important paradigm for enhancing the reasoning capabilities of large language models (LLMs) in recent years, with applications spanning machine translation~\citep{liang2024encouraging}, retrieval-augmented generation~\citep{chang2025main}, and LLM-based evaluation~\citep{chan2024chateval}. The standard pipeline has multiple LLM agents engage in multi-round structured discussions on the same problem and aggregates their answers via Majority Voting~(MV)~\citep{du2024improving}, extending the majority-vote paradigm introduced by Self-Consistency~\citep{wang2023selfconsistency} from single-model sampling to multi-agent interaction. However, the effectiveness of this paradigm rests on a critical assumption from the Condorcet Jury Theorem: the errors of individual voters must be mutually independent.

Yet contemporary LLMs share highly similar pretraining corpora, training procedures, and even architectural designs, causing their errors to exhibit strong correlations that fundamentally violate this independence assumption~\citep{kim2025correlated}. Recent studies have confirmed this concern from multiple angles: correlated errors can lock in incorrect answers through a ``Tyranny of the Majority'' effect~\citep{estornell2024multi}, agents rarely correct stance biases during debate~\citep{wu2025debate}, and debate itself may not improve reasoning correctness when the aggregation strategy is suboptimal~\citep{choi2025debate}. In other words, the problem is not ``how to argue'' but ``how to count.''

These findings collectively point to a systematically overlooked phenomenon: when opinion divergence arises in multi-agent debates, the minority holding the correct answer is suppressed by the majority voting mechanism. We term this phenomenon \emph{Minority Truth}, drawing on Moscovici's minority influence theory in social psychology~\citep{moscovici1976social}, which showed that consistent minorities can challenge and even overturn majority consensus. In the LLM debate setting, we ask the question: when does the minority already hold the correct answer, and how can we detect this?

Through structured debates among three heterogeneous LLM agents\footnote{For ease of presentation, we assume three heterogeneous agents in this paper. However, our approach applies naturally to cases with a larger number of agents, which we leave for future work.} from different vendors on six public benchmarks, we find this phenomenon to be remarkably prevalent. Among all 1,754 questions, 39.1\% of samples (686/1,754) exhibit 2:1 opinion divergences, and in 25.5\% of these divergent samples (175/686) the minority actually holds the correct answer. This means that majority voting suppresses the correct answer in approximately one out of every four disagreements, corresponding to a theoretical recovery margin of 10.0 percentage points between Majority Voting accuracy (74.3\%) and the Oracle upper bound (84.3\%).

Faced with this problem, a natural recourse is to employ another, more powerful LLM as the final arbiter. The LLM-as-Judge paradigm has been widely adopted for evaluation tasks~\citep{zheng2023judging}, and recent work has explored multi-agent debate among LLM judges to improve judgment robustness~\citep{hu2025multi}. However, when applied to overturning majority voting in MAD, these approaches inherit a fundamental limitation: the judge LLM shares similar knowledge blind spots with the debating agents~\citep{kim2025correlated}. Our experiments confirm this concern: an LLM judge reading debate logs and making rulings yields negative Net Gain, performing worse than no intervention at all (detailed in Section~\ref{sec:setup}). When the majority errs due to shared reasoning biases, the judge is highly likely to make the same mistake. Therefore, we need an approach that breaks out of the LLM cognitive closed loop.

Our key insight is that while LLMs' answers may be correlated due to shared training data, their behavioral patterns during debate, such as how readily agents change positions, whether they introduce genuinely new arguments, and how reasoning quality distributes across sides, carry discriminative signals that a non-LLM classifier can exploit.

To this end, we propose \textbf{\MS{}}, a lightweight, plug-and-play meta-classifier framework that employs a \emph{Diagnosis--Cure} dual-phase paradigm. In the Diagnosis phase, the system collects divergent samples from multi-agent debates. In the Cure phase, it extracts behavioral features from debate logs and trains a lightweight non-LLM meta-classifier to learn ``when to overturn majority voting.'' By deliberately choosing a non-LLM classifier, the framework achieves \emph{cognitive orthogonality} with the debating agents, as its judgment basis (statistical behavioral patterns) is fundamentally different from the semantic reasoning that produced the correlated errors in the first place. The framework requires no modification to any underlying LLM's weights or inference pipeline; it merely inserts a lightweight ``safety valve'' at the voting aggregation layer.

The main contributions of this paper are as follows:

\begin{itemize} [leftmargin=*]
\item We conduct a systematic quantitative analysis of the \emph{Minority Truth} phenomenon in multi-agent debates, revealing its prevalence across six benchmarks: 25.5\% of divergent cases have the minority holding the correct answer, yielding a 10.0-percentage-point recovery margin.
\item We propose the \emph{Diagnosis--Cure} framework: a plug-and-play paradigm that operates entirely at the aggregation layer, requiring no modification to any underlying LLM's weights, prompts, or inference pipeline.
\item We design a multi-granularity behavioral feature system, the \emph{debate fingerprint}, that captures \emph{how agents argued} (debate dynamics), \emph{what the voting structure looks like} (voting metadata), and \emph{who made the stronger case} (semantic audit).
\item We rigorously validate the framework, achieving Flip Precision of 81.2\% with positive Net Gain on all six datasets and all 20 random seed trials, while the LLM-as-Judge baseline yields negative gains, confirming that behavioral-feature-based meta-classification outperforms LLM adjudication for minority recovery.
\end{itemize}

The remainder of the paper is organized as follows: Section~2 reviews related work, Section~3 details the method, Section~4 describes the experimental setup and presents results, Section~5 provides in-depth analysis, Section~6 discusses implications and limitations, and Section~7 concludes.

\section{Related Work}

\subsection{Multi-Agent Debate and Conformity Effects}

Leveraging interactions among multiple LLM agents to improve reasoning quality has become an important research direction. \citet{du2024improving} proposed the multi-agent debate framework, having multiple LLMs engage in multi-round discussions and demonstrating that the debate process effectively improves factuality and reasoning. This work inspired extensive follow-up research applying multi-agent debate to mathematical reasoning, code generation, commonsense QA, and other scenarios. However, conformity effects during debate have gradually attracted attention. \citet{wu2025debate} found through quantitative experiments that weak models correct only 3.6\% of stance biases during debates, indicating that agents tend to abandon their correct judgments and conform to the majority. \citet{choi2025debate} further pointed out that debate itself does not improve reasoning correctness: when controlling for the aggregation strategy, debate and independent answering show no significant difference in accuracy, and the real bottleneck lies in extracting the correct answer from disagreements. These findings reveal a core contradiction: debate may expose errors through the clash of perspectives, but it may also suppress correct minority opinions through social pressure, and existing work lacks mechanisms to distinguish between these two scenarios.

\subsection{Failure of Majority Voting and Alternative Aggregation}

As the simplest aggregation strategy, majority voting has its theoretical foundation in the Condorcet Jury Theorem. \citet{wang2023selfconsistency} combined majority voting with Chain-of-Thought reasoning to propose Self-Consistency, achieving significant improvements on multiple reasoning tasks. However, when voter errors are no longer independent, the theoretical guarantee of majority voting breaks down. \citet{estornell2024multi} provided a formal analysis, proving that under highly correlated LLM errors, majority voting can systematically lock in incorrect answers, a phenomenon they termed ``Tyranny of the Majority.'' To address these limitations, researchers have explored weighted aggregation using second-order information such as consistency structures among agents~\citep{ai2025beyond}. However, most of these methods operate at the voting level and have not fully exploited the rich behavioral signals embedded in the debate process. The \emph{debate fingerprint} feature system proposed in this paper targets precisely this gap: we attend not only to ``who voted for what'' but also to ``what happened during the debate.''

\subsection{Debate Quality Auditing and Minority Recovery}

Recent work has begun to directly address debate quality auditing and minority opinion recovery. \citet{yang2026agentauditor} proposed AgentAuditor, which constructs reasoning trees and performs structured auditing to assess multi-agent system output quality, representing the most closely related work. AgentAuditor replaces majority voting with evidence-based path search over a Reasoning Tree, auditing only decision-critical divergence points. However, AgentAuditor focuses on post-hoc auditing of all agent outputs rather than selective intervention specifically in debate divergence scenarios, and its Anti-Consensus Preference Optimization requires constructing training data from historical majority-failure cases.

Compared to AgentAuditor and other recent approaches, \MS{} differentiates itself in three aspects. First, we focus on debate divergence, framing the problem as a binary decision of ``whether to flip the majority vote'' rather than general output quality assessment. Second, our \emph{debate fingerprint} fuses debate dynamics, voting metadata, and semantic audit signals, extracting directly from debate logs without additional reasoning tree construction. Third, we employ LightGBM rather than neural networks or LLMs as the meta-classifier, maintaining lightweight efficiency and interpretability while circumventing the systematic interference of LLM error correlation on adjudication accuracy.

\section{Our Method}

\begin{figure*}[!t]
\centering
\includegraphics[width=\textwidth, trim=12 67 8 20, clip]{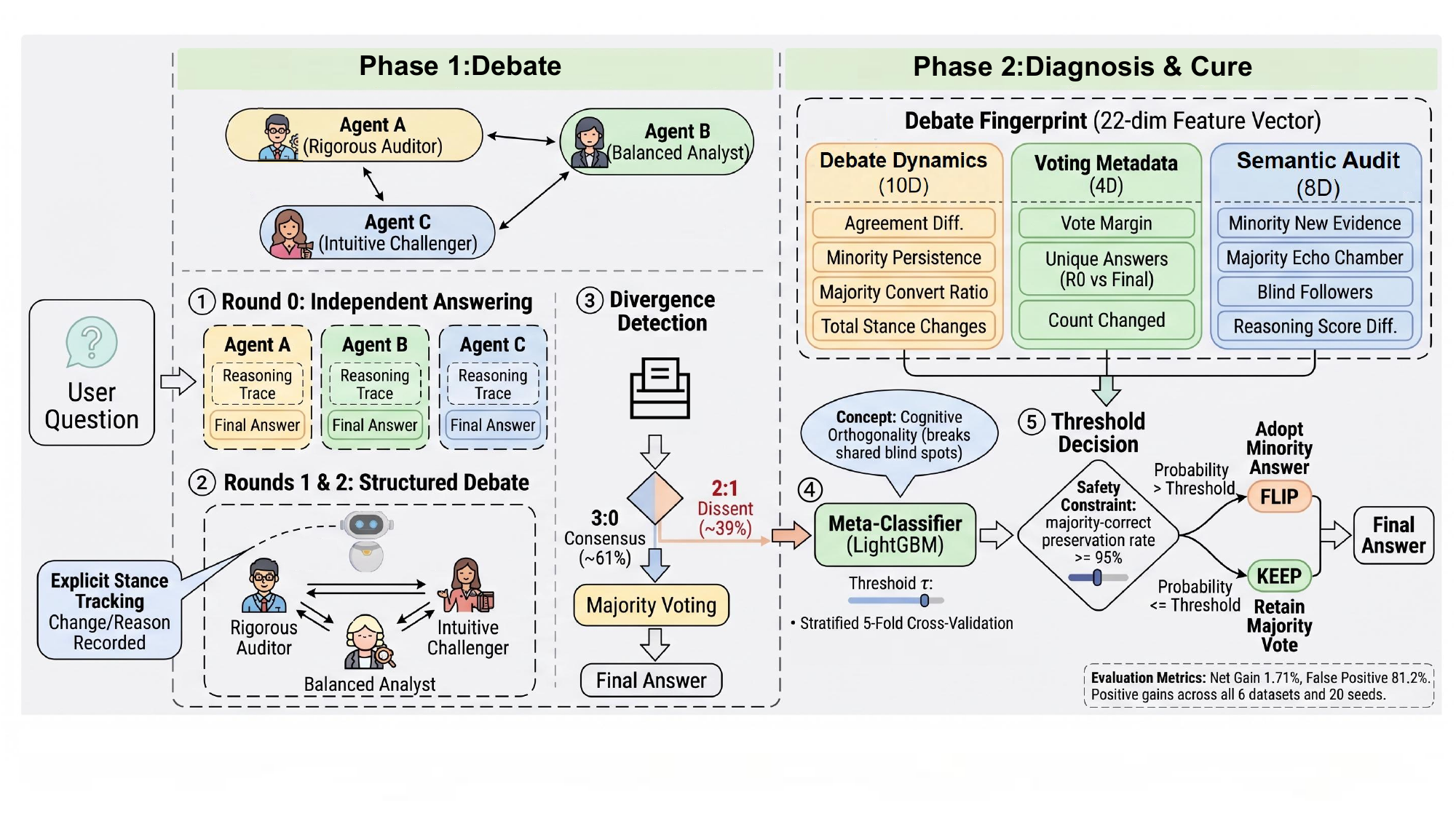}
\caption{System overview of the \MS{} framework. The Diagnosis phase collects divergent samples through multi-agent debate; the Cure phase extracts a multi-dimensional \emph{debate fingerprint} and trains a LightGBM meta-classifier to determine when to overturn majority voting.}
\Description{A flowchart showing three phases: six benchmarks feed into a multi-agent debate with three LLM agents (GPT-4o-mini, Gemini-2.0-Flash, Claude Haiku 4.5), producing divergent samples that are processed through a multi-dimensional debate fingerprint (10 dynamics, 4 meta, 8 semantic audit features) and a LightGBM classifier with threshold-based flip decision.}
\label{fig:system_overview}
\end{figure*}

In this section, we present the \MS{} framework. We first formulate the problem (\S3.1), then describe the Diagnosis phase for collecting divergent samples (\S3.2), the debate fingerprint feature system (\S3.3), and the Cure phase for meta-classification and threshold optimization (\S3.4). Figure~\ref{fig:system_overview} provides an overview: the system operates in two phases, Diagnosis (multi-agent debate and divergence detection) and Cure (feature extraction, classification, and flip decision).

\subsection{Problem Formulation}

\paragraph{Multi-Agent Debate (MAD)}
Consider a system of $K$ LLM agents. Given a question $q$, each agent produces a final answer $a_k$ ($k = 1, \ldots, K$) after $R$ rounds of structured debate. Majority voting selects the answer receiving the most votes as the system output $\hat{a}_{\text{MV}}$.

\paragraph{Divergent Sample.}
When the voting outcome exhibits divergence (i.e., at least one agent's final answer differs from the majority), we call this a \emph{divergent sample}. In our setting ($K = 3$), this corresponds to a 2:1 split.

\paragraph{Minority Truth.}
Among divergent samples, if the minority's answer matches the ground truth while the majority's answer is incorrect, we term this \emph{Minority Truth}. The prevalence of Minority Truth directly determines the theoretical recovery ceiling.

\paragraph{Evaluation Metrics.}
The core constraint of selective intervention is that each Wrong Flip (WF) incurs a cost exactly equal to the benefit of a Correct Flip (CF). We adopt Net Gain (NG) as the primary metric, formulated as:
\begin{equation}
\label{eq:net_gain}
\text{NG} = \frac{\text{CF} - \text{WF}}{N_{\text{total}}}
\end{equation}
where $N_{\text{total}}$ is the total number of questions (including non-divergent samples). $\text{NG} > 0$ indicates net benefit; $\text{NG} < 0$ means intervention is harmful. We additionally define Flip Precision $\text{FP} = \text{CF} / (\text{CF} + \text{WF})$ and Recovery Rate $\text{CF} / N_{\text{MT}}$, where $N_{\text{MT}}$ is the number of Minority Truth samples.

\paragraph{Problem Definition.}
Our goal is to learn a decision function $f(\mathbf{x})$ on divergent samples, where $\mathbf{x}$ is the feature vector extracted from the debate log: $f(\mathbf{x}) = 1$ triggers a flip to the minority answer; $f(\mathbf{x}) = 0$ preserves the majority vote. For non-divergent samples, the system always retains the majority voting result.

\subsection{Diagnosis Phase: Multi-Agent Debate Protocol}

The Diagnosis phase aims to systematically collect divergent samples and their debate logs. We construct a debate system comprising $K$ heterogeneous LLM agents, each satisfying two diversity requirements: (1)~\emph{architectural diversity}: agents are sourced from different model vendors with distinct pretraining corpora and architectures, increasing the likelihood that their knowledge blind spots do not fully overlap; and (2)~\emph{cognitive role diversity}: each agent is assigned a distinct reasoning persona via system prompt, such as a conservative auditor who resists conformity, a balanced analyst who weighs evidence fairly, or an intuitive challenger who explores unconventional paths. All agents share the same temperature to isolate the effect of model and role diversity from sampling randomness. The specific model choices and role assignments used in our experiments are detailed in Section~\ref{sec:setup}.

The debate protocol is designed around two principles. \emph{Independence before interaction}: agents first answer independently (Round~0) before seeing others' views, ensuring unbiased initial opinions and providing a clean baseline for measuring subsequent opinion shifts. \emph{Explicit stance tracking}: during debate rounds, agents must explicitly declare whether they changed position and why, generating the structured behavioral signals that the Cure phase later relies on for feature extraction.

Concretely, the debate proceeds in three phases. In Round~0 (independent answering), each agent independently reads the question and provides an initial answer with reasoning; agents cannot see each other's responses. In Rounds~1 and~2 (debate), each agent sees all other agents' responses from the previous round, updates their answer, and explicitly indicates whether and why they changed their position. After debate concludes, the system performs majority voting on the final-round answers and passes all samples with 2:1 divergence and their complete debate logs to the Cure phase.

The choice of three agents and two debate rounds represents a principled trade-off between signal richness and noise control. Three agents produce the simplest non-trivial divergence structure (2:1), which yields a clean binary classification target: minority correct vs.\ majority correct. With five or more agents, divergence patterns become combinatorially complex (e.g., 3:2, 4:1, or multi-way splits), each requiring separate modeling and fragmenting the already limited training data. Furthermore, under the error correlation documented by~\citet{kim2025correlated}, adding more agents of similar capability may amplify rather than attenuate systematic biases. Two debate rounds provide sufficient stance dynamics information while avoiding excessive opinion convergence that could eliminate the divergence signal. We leave exploration of larger agent panels and deeper debate to future work (Section~\ref{sec:discussion}).

\subsection{The Debate Fingerprint}

The core design philosophy of \MS{} is that debate logs contain rich behavioral signals sufficient to determine whether the majority consensus is reliable. We extract a multi-dimensional feature vector (the \emph{debate fingerprint}), organized into three complementary groups comprising 22 features in total (Table~\ref{tab:features}).

\textbf{Debate Dynamics.} (10 dim.) directly characterize agent behavioral patterns. The design intuition: if the majority internally endorses each other without independent argumentation, while the minority consistently maintains its position and introduces new evidence, the majority consensus warrants suspicion.

\textbf{Voting Metadata.} (4 dim.) captures the macro structure of opinion evolution, e.g., three distinct answers in Round~0 that converge to two, at least one agent was persuaded to change position.

\textbf{Semantic Audit.} (8 dim.) invokes GPT-4o (temperature~=~0.0) as an external logic auditor to assess reasoning quality without knowledge of the correct answer, compensating for semantic information that purely statistical features cannot capture.

The three groups are functionally complementary: Dynamics captures ``how they argued,'' Meta captures ``what the voting structure looks like,'' and Audit captures ``who made the stronger case.''

\begin{table}[!htbp]
\centering
\caption{The \emph{debate fingerprint}, organized by signal source.}
\label{tab:features}
\scriptsize
\setlength{\tabcolsep}{3pt}
\renewcommand{\arraystretch}{0.94}
\begin{tabular}{@{}p{0.29\columnwidth}p{0.64\columnwidth}@{}}
\toprule
\textbf{Feature} & \textbf{Description} \\

\midrule
\multicolumn{2}{l}{\emph{Debate Dynamics (10 dim.)---how agents argued}} \\[2pt]
total\_stance\_changes
  & Total position changes across all agents and rounds \\
majority\_convert\_ratio
  & Fraction of majority camp that switched from minority \\
minority\_persistence
  & Whether minority maintained position across all rounds \\
minority\_new\_info\_ratio
  & Ratio of rounds where minority introduced new arguments \\
majority\_new\_info\_ratio
  & Ratio of rounds where majority introduced new arguments \\
majority\_agreement\_count
  & Endorsements received by majority reasoning \\
minority\_agreement\_count
  & Endorsements received by minority reasoning \\
agreement\_diff
  & majority\_agreement\_count $-$ minority\_agreement\_count \\
explicit\_changes\_majority
  & Explicit stance changes within majority camp \\
explicit\_changes\_minority
  & Explicit stance changes within minority camp \\

\midrule
\multicolumn{2}{l}{\emph{Voting Metadata (4 dim.)---voting structure}} \\[2pt]
vote\_margin
  & Vote count difference (majority $-$ minority) \\
num\_unique\_answers\_r0
  & Number of distinct answers in Round~0 \\
num\_unique\_answers\_final
  & Number of distinct answers in final round \\
answers\_changed
  & Number of agents whose answer changed from R0 to final \\

\midrule
\multicolumn{2}{l}{\emph{Semantic Audit (8 dim.)---reasoning quality}} \\[2pt]
minority\_new\_evidence
  & Whether minority introduces new evidence (0/1) \\
majority\_echo\_chamber
  & Whether majority exhibits echo chamber effect (0/1) \\
minority\_finds\_error
  & Whether minority identifies logical error in majority (0/1) \\
majority\_logical\_gap
  & Whether majority argumentation has logical gaps (0/1) \\
minority\_reasoning\_score
  & GPT-4o quality score for minority reasoning (1--5) \\
majority\_reasoning\_score
  & GPT-4o quality score for majority reasoning (1--5) \\
reasoning\_score\_diff
  & minority $-$ majority reasoning score (computed client-side) \\
blind\_follower\_count
  & Agents who switched without substantive reasoning (0--2) \\

\bottomrule
\end{tabular}
\end{table}

\subsection{Cure Phase: Meta-Classifier and Threshold Strategy}

In the Cure phase, we employ LightGBM~\citep{ke2017lightgbm} as the meta-classifier, taking the \emph{debate fingerprint} as input and outputting a continuous probability $P(\text{minority correct})$. The choice of LightGBM is motivated by three reasons. First, gradient boosting trees excel at handling small samples (686 divergent samples), heterogeneous features (Boolean, count, and continuous types), and nonlinear feature interactions. Second, tree-based feature importance provides natural interpretability. Third, and most critically, a non-LLM classifier fundamentally circumvents the systematic interference of LLM error correlation on adjudication accuracy, as our experiments demonstrate: directly using an LLM as judge leads to negative gains.

A critical design challenge is balancing recall (recovering as many Minority Truth cases as possible) against safety (avoiding wrong flips that negate gains). Since each wrong flip exactly cancels one correct flip in the Net Gain metric, an overly aggressive strategy (low threshold) easily produces negative returns, as demonstrated by the LLM-as-Judge baseline. To navigate this recall--safety trade-off, the continuous probability output is converted to a binary decision via threshold $\tau$: when $P > \tau$, the system flips; otherwise the majority vote is preserved.

We adopt a \emph{per-dataset threshold optimization} strategy, performing a grid search over $\tau \in [0.05, 0.95]$ with step size 0.01 on each dataset's out-of-fold predictions, independently selecting the $\tau$ that maximizes Net Gain subject to the safety constraint that the majority-correct preservation rate remains $\geq$\,95\%, i.e., at most 5\% of originally correct majority decisions may be overturned. This conservative threshold reflects a ``first, do no harm'' design principle: in a system where the majority is already correct in 74.5\% of divergent cases, flip safety must take precedence over recall. This adaptive, per-dataset strategy allows the model to adjust decision aggressiveness for domains with different Minority Truth prevalence rates.

To prevent train--test leakage, we employ Stratified 5-Fold Cross-Validation with labels as the stratification criterion. All reported metrics are aggregated from out-of-fold predictions across five folds. With the framework fully specified, we now describe the experimental setup and present results.

\section{Experiments and Results}
\label{sec:setup}

\subsection{Datasets and Debate Configuration}

We select six public benchmark datasets covering diverse reasoning capabilities: ARC-Challenge~\citep{clark2018arc} tests scientific commonsense reasoning, CommonsenseQA~\citep{talmor2019commonsenseqa} tests commonsense reasoning, GSM8K~\citep{cobbe2021gsm8k} tests mathematical reasoning, MMLU-STEM~\citep{hendrycks2021mmlu} covers multi-discipline STEM knowledge, TruthfulQA~\citep{lin2022truthfulqa} tests factual judgment and deception resistance, and WinoGrande~\citep{sakaguchi2021winogrande} tests pronoun disambiguation.

The three debate agents are instantiated as follows. Agent~A uses OpenAI GPT-4o-mini with the role of ``Rigorous Auditor,'' emphasizing logical verification and resistance to conformity. Agent~B uses Google Gemini-2.0-Flash with the role of ``Balanced Analyst,'' favoring fair evaluation of all arguments. Agent~C uses Anthropic Claude Haiku~4.5 with the role of ``Intuitive Challenger,'' encouraging creative thinking but relatively susceptible to confident-sounding arguments. All three agents use temperature~=~0.7; cognitive diversity arises entirely from different vendor models and distinct role assignments rather than temperature variation. The verbatim system prompts are provided in Appendix~\ref{app:prompts}.

After Round~0 independent answering and two rounds of debate, a total of 1,754 deduplicated debate logs are produced. Among the 1,754 debate logs, 686 (39.1\%) exhibit 2:1 opinion divergence, constituting the core analytical subjects of this study. Among these 686 divergent samples, 175 (25.5\%) have the minority holding the correct answer. This means the Majority Voting accuracy is 74.3\%, while perfect identification of all Minority Truth cases would yield 84.3\% (Oracle Upper Bound); the 10.0\% gap represents the theoretical recovery ceiling. Significant heterogeneity exists across datasets: CommonsenseQA has the highest Minority Truth rate (47.0\%), while WinoGrande has the lowest (14.0\%). Table~\ref{tab:dataset_stats} summarizes the detailed statistics.

\begin{table}[!htbp]
\centering
\caption{Dataset statistics across six benchmarks. $N$ = total questions, $N_{\text{div}}$ = divergent samples, $N_{\text{MT}}$ = Minority Truth count, MT\% = Minority Truth rate, $\Delta_{\max}$ = theoretical recovery margin.}
\label{tab:dataset_stats}
\footnotesize
\begin{tabular}{@{}lrrrrrrr@{}}
\toprule
\textbf{Dataset} & $N$ & $N_{\text{div}}$ & $N_{\text{MT}}$ & \textbf{MT\%} & \textbf{MV\%} & \textbf{Orac.\%} & $\Delta_{\max}$ \\
\midrule
ARC-Chall.     & 96  & 33  & 14 & 42.4 & 71.9 & 86.5 & 14.6 \\
CSQA           & 253 & 115 & 54 & 47.0 & 54.9 & 76.3 & 21.3 \\
GSM8K          & 137 & 76  & 16 & 21.1 & 70.8 & 82.5 & 11.7 \\
MMLU-STEM      & 726 & 236 & 49 & 20.8 & 80.9 & 87.6 & 6.7  \\
TruthfulQA     & 210 & 76  & 21 & 27.6 & 62.4 & 72.4 & 10.0 \\
WinoGrande     & 332 & 150 & 21 & 14.0 & 84.6 & 91.0 & 6.3  \\
\midrule
\textbf{Total} & \textbf{1754} & \textbf{686} & \textbf{175} & \textbf{25.5} & \textbf{74.3} & \textbf{84.3} & \textbf{10.0} \\
\bottomrule
\end{tabular}
\end{table}

\paragraph{Evaluation Protocol.}
As defined in Eq.~(\ref{eq:net_gain}), we use Net Gain (NG) as the primary evaluation metric, with Flip Precision (FP), Recovery Rate, and AUC as auxiliary metrics. We employ Stratified 5-Fold Cross-Validation with labels as stratification criterion. Threshold optimization is performed independently on each dataset, searching for the optimal $\tau$ maximizing NG subject to the constraint that the majority-correct preservation rate remains $\geq$\,95\%. All metrics are aggregated from out-of-fold predictions.

\paragraph{Baseline Methods.}
We compare \MS{} against five baselines. \textbf{Majority Voting (MV)} is the reference with NG defined as 0\%. \textbf{Always Trust Minority} unconditionally flips all divergent samples. \textbf{Single Best Feature} uses only the most discriminative single feature (\texttt{reasoning\_score\_diff}) for threshold classification. \textbf{Logistic Regression} uses the same \emph{debate fingerprint} features and CV framework but replaces LightGBM with logistic regression. \textbf{LLM-as-Judge} uses GPT-4o (temperature~=~0.0) to directly read complete debate logs and make rulings.

\subsection{Per-Dataset Performance}

Under the multi-dimensional \emph{debate fingerprint} and Stratified 5-Fold CV framework, \MS{} achieves positive Net Gain on all six datasets, with an overall NG~=~+1.71\%, corresponding to 39 Correct Flips and 9 Wrong Flips, Flip Precision of 81.2\%, and Recovery Rate of 22.3\% (39/175).

Per-dataset results exhibit differentiation closely related to task characteristics. \textbf{GSM8K} stands out most prominently, with AUC reaching 0.957, NG~=~+8.03\%, and all 11 flips correct (WF~=~0). This is primarily due to the high discriminability of debate signals in mathematical reasoning: behavioral patterns such as the minority introducing new computation steps or pointing out the majority's arithmetic errors are extremely clear. \textbf{ARC-Challenge} also performs excellently (NG~=~+5.21\%), with all 5 flips correct.

\textbf{CommonsenseQA} and \textbf{MMLU-STEM}, as the two datasets with the most divergent samples (115 and 236), are also the only two datasets that incur wrong flips (2 and 7, accounting for all 9 WFs); MMLU-STEM alone contributes 16 of the 39 correct flips. CommonsenseQA's Flip Precision of 60.0\% is the lowest, consistent with commonsense reasoning where multiple answers possess some plausibility. \textbf{TruthfulQA} achieves a robust +1.43\% gain (CF~=~3, WF~=~0). \textbf{WinoGrande} shows the smallest gain (NG~=~+0.30\%), executing only 1 correct flip, consistent with its extremely low Minority Truth rate (14.0\%) and the correspondingly conservative threshold ($\tau$~=~0.94).

Notably, no dataset shows negative NG, indicating that Sentinel achieves at least ``do no harm'' across all domains. Overall, from the 10.0\% theoretical recovery margin, Sentinel recovers 1.71 percentage points (17.1\% of the theoretical ceiling). Table~\ref{tab:main_results} presents the complete per-dataset results.

\begin{table}[!htbp]
\centering
\caption{Main experimental results per dataset. $\tau$ = optimized threshold. All datasets achieve positive Net Gain.}
\label{tab:main_results}
\footnotesize
\begin{tabular}{@{}lrrrrrrr@{}}
\toprule
\textbf{Dataset} & $\tau$ & \textbf{AUC} & \textbf{CF} & \textbf{WF} & \textbf{FP\%} & \textbf{Rec\%} & \textbf{NG\%} \\
\midrule
ARC-Chall.    & .85 & .823 & 5  & 0 & 100.0 & 35.7 & +5.21 \\
CSQA          & .88 & .612 & 3  & 2 & 60.0  & 5.6  & +0.40 \\
GSM8K         & .60 & .957 & 11 & 0 & 100.0 & 68.8 & +8.03 \\
MMLU-STEM     & .76 & .746 & 16 & 7 & 69.6  & 32.7 & +1.24 \\
TruthfulQA    & .87 & .681 & 3  & 0 & 100.0 & 14.3 & +1.43 \\
WinoGrande    & .94 & .581 & 1  & 0 & 100.0 & 4.8  & +0.30 \\
\midrule
\textbf{Overall} & — & \textbf{.741} & \textbf{39} & \textbf{9} & \textbf{81.2} & \textbf{22.3} & \textbf{+1.71} \\
\bottomrule
\end{tabular}
\end{table}

\subsection{Baseline Comparison}

To assess the advantages of \MS{}, we compare it against five baselines, all evaluated on the same 686 divergent samples with $N_{\text{total}} = 1{,}754$ as the NG denominator.

\textbf{Always Trust Minority} yields NG~=~$-$19.16\%, demonstrating that blindly trusting the minority is catastrophic (CF~=~175, WF~=~511).
\textbf{Single Best Feature} achieves NG~=~+0.23\% (CF~=~7, WF~=~3), demonstrating the necessity of multi-feature fusion.
\textbf{Logistic Regression} achieves NG~=~+0.68\%, validating the \emph{debate fingerprint} while showing that LightGBM's nonlinear modeling provides an additional +1.03\% gain.

The most analytically valuable baseline is \textbf{LLM-as-Judge}. GPT-4o directly reading debate logs yields NG~=~$-$1.37\% (CF~=~70, WF~=~94), worse than no intervention. Its Recovery Rate reaches 40.0\%, far exceeding Sentinel's 22.3\%, but Flip Precision of only 42.7\% means fewer than half its flips are correct. GPT-4o shares similar knowledge blind spots with the debating agents, corroborating our core premise: LLM errors are highly correlated, and another LLM as arbiter cannot solve the problem. Table~\ref{tab:baseline} presents the full comparison.

\begin{table}[!htbp]
\centering
\caption{Baseline comparison. \MS{} is the only method achieving both high Flip Precision and positive Net Gain.}
\label{tab:baseline}
\footnotesize
\begin{tabular}{@{}lrrrrr@{}}
\toprule
\textbf{Method} & \textbf{CF} & \textbf{WF} & \textbf{FP\%} & \textbf{Rec\%} & \textbf{NG\%} \\
\midrule
Majority Voting    & --  & --  & --   & --    & 0.00  \\
Always Trust Min.  & 175 & 511 & 25.5 & 100.0 & $-$19.16 \\
Single Best Feat.  & 7   & 3   & 70.0 & 4.0   & +0.23 \\
Logistic Regr.     & 24  & 12  & 66.7 & 13.7  & +0.68 \\
LLM-as-Judge       & 70  & 94  & 42.7 & 40.0  & $-$1.37 \\
\textbf{Sentinel}  & \textbf{39} & \textbf{9} & \textbf{81.2} & \textbf{22.3} & \textbf{+1.71} \\
\midrule
Oracle (upper bd.) & 175 & 0   & 100  & 100   & +9.98 \\
\bottomrule
\end{tabular}
\end{table}

\section{Ablation, Interpretability, and Robustness}

\subsection{Ablation Study}

To verify each feature group's contribution, we conduct two sets of ablation experiments: ``remove one group'' and ``keep only one group,'' all using identical LightGBM hyperparameters, Stratified 5-Fold CV, and per-dataset threshold optimization.

Removing Dynamics decreases NG from +1.71\% to +1.14\% ($\Delta=-0.57\%$), the largest impact among three groups, with WF increasing from 9 to 15 and FP dropping from 81.2\% to 70.0\%, indicating that debate dynamics features are the key barrier controlling wrong flips. Removing Semantic Audit decreases NG to +1.43\% ($\Delta$~=~$-$0.29\%), with CF dropping from 39 to 32, showing substantial incremental recall. Removing Meta slightly increases NG to +1.77\% ($\Delta$~=~+0.06\%), but this comes with increased aggressiveness (CF rises from 39 to 42, WF from 9 to 11). The marginal gain is statistically insignificant and likely reflects threshold re-optimization on the altered probability surface rather than true feature redundancy; in other settings, the higher WF rate could erode net gains.

Single-group experiments corroborate these conclusions. Dynamics alone achieves NG~=~+1.25\% (CF~=~31, WF~=~9), recovering 73\% of the full model's net gain, demonstrating its dominant ``backbone'' position. Semantic Audit alone yields NG~=~+0.57\% (CF~=~17, WF~=~7). Meta alone yields NG~=~+0.51\% (AUC~=~0.585), the weakest discriminative ability. Table~\ref{tab:ablation} presents the complete results.

\begin{table}[!htbp]
\centering
\caption{Ablation study. Top: removing one feature group. Bottom: keeping only one group. Dynamics is the backbone; Semantic Audit provides precise incremental gains.}
\label{tab:ablation}
\footnotesize
\begin{tabular}{@{}lcrrrrr@{}}
\toprule
\textbf{Configuration} & \textbf{\#F} & \textbf{AUC} & \textbf{CF} & \textbf{WF} & \textbf{NG\%} & $\Delta$ \\
\midrule
\multicolumn{7}{l}{\emph{Remove one group}} \\
Full (22 features)     & 22 & .741 & 39 & 9  & +1.71 & —     \\
\; $-$ Dynamics           & 12 & .712 & 35 & 15 & +1.14 & $-$0.57 \\
\; $-$ Meta               & 18 & .732 & 42 & 11 & +1.77 & +0.06 \\
\; $-$ Semantic Audit          & 14 & .729 & 32 & 7  & +1.43 & $-$0.29 \\
\midrule
\multicolumn{7}{l}{\emph{Keep only one group}} \\
Dynamics Only          & 10 & .722 & 31 & 9 & +1.25 & — \\
Meta Only              & 4  & .585 & 17 & 8 & +0.51 & — \\
Semantic Audit Only         & 8  & .680 & 17 & 7 & +0.57 & — \\
\bottomrule
\end{tabular}
\end{table}

\subsection{Feature Importance}

To understand \MS{}'s decision basis, we train LightGBM on all 686 divergent samples and extract split importance (total split count across all trees).

The top three features all come from the Dynamics group: \feat{agreement\_diff} (split\,=\,234), \feat{minority\_agreement\_count} (split\,=\,233), and \feat{majority\_agreement\_count} (split\,=\,226). Together, they characterize the agreement structure during debate. The intuition is clear: if substantive endorsement received by the minority approaches that of the majority, the majority's advantage is not robust and flipping has higher value.

\begin{figure}[!htbp]
\centering
\includegraphics[width=0.92\columnwidth]{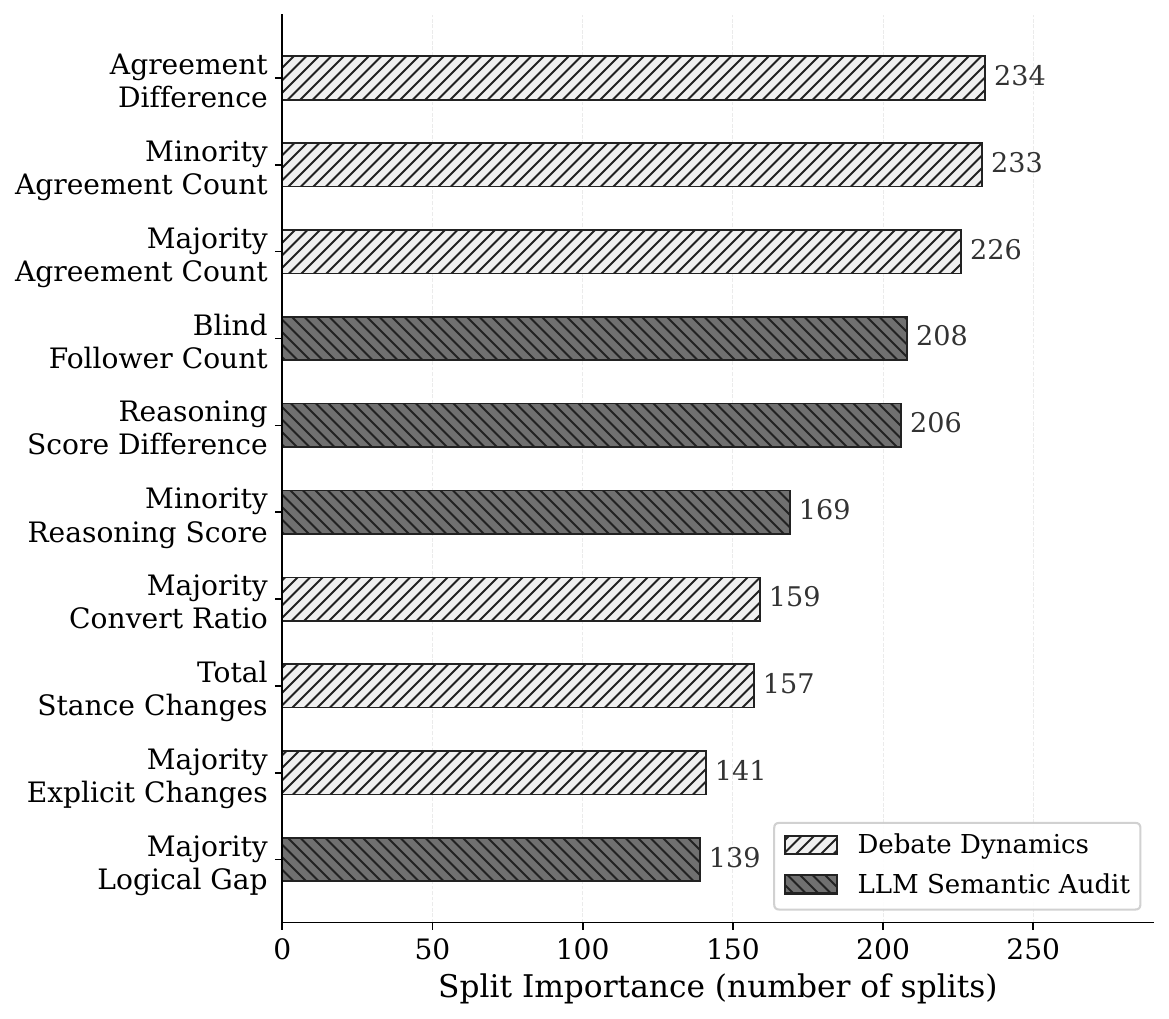}
\caption{Top-10 feature importance by split count. The top three are all Dynamics features characterizing agreement structure.}
\Description{Horizontal bar chart showing top-10 features ranked by LightGBM split count. Agreement Diff leads at 234 splits, followed by Minority Agreement Count (233) and Majority Agreement Count (226). Six bars are colored blue for Dynamics features and four orange for Semantic Audit features.}
\label{fig:feature_importance}
\end{figure}

Following closely are two Semantic Audit features: \texttt{blind\_\allowbreak follower\_\allowbreak count} (split\,=\,208) and \texttt{reasoning\_\allowbreak score\_\allowbreak diff} (split\,=\,206). The former detects blind followers in the majority camp; the latter measures the reasoning quality score difference. Among the Top~10, Dynamics occupies 6 positions, Semantic Audit occupies 4, and Meta has zero entries, highly consistent with ablation conclusions. Figure~\ref{fig:feature_importance} presents the Top-10 ranking.

\subsection{Threshold Robustness}

The threshold $\tau$ is the sole parameter converting probability output into binary decisions. We perform a global fixed threshold sweep from 0.05 to 0.95 (step 0.01) on out-of-fold prediction probabilities.

\begin{figure*}[!tp]
\centering
\begin{subfigure}[t]{0.48\textwidth}
\centering
\includegraphics[width=\linewidth]{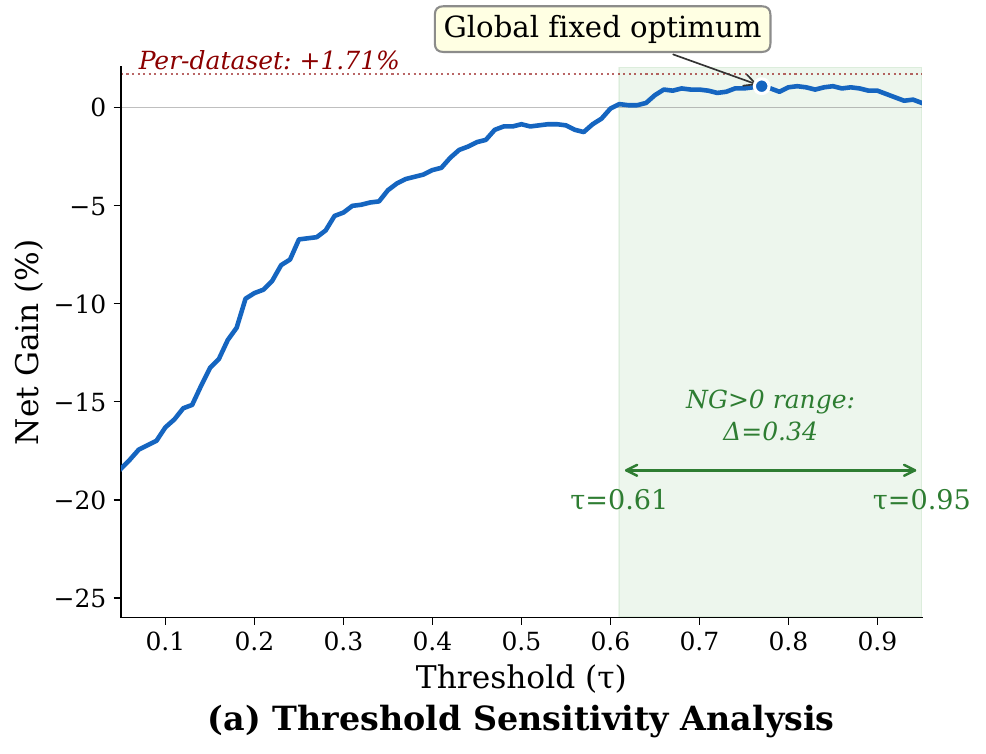}
\caption{Threshold sensitivity. Net Gain remains positive across $\tau \in [0.61, 0.95]$ (width 0.34). The globally optimal fixed threshold is $\tau = 0.81$ yielding NG~=~+1.08\%; the dashed red line marks the per-dataset result (+1.71\%) for reference.}
\label{fig:threshold}
\end{subfigure}
\hfill
\begin{subfigure}[t]{0.48\textwidth}
\centering
\includegraphics[width=\linewidth]{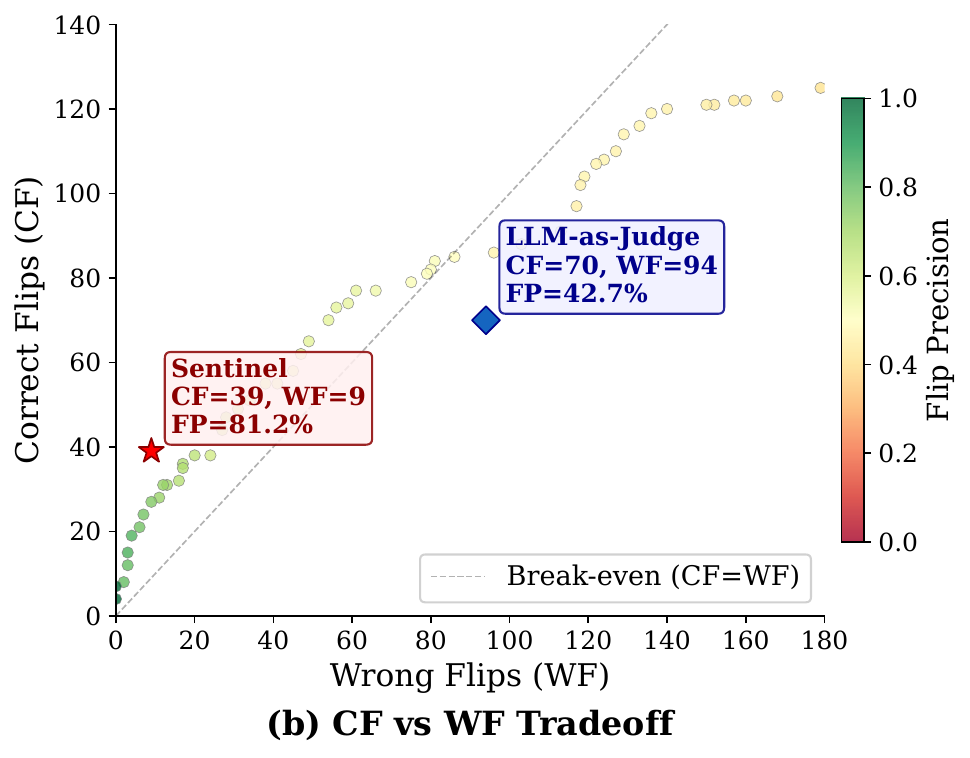}
\caption{CF vs.\ WF trade-off along the threshold sweep. Sentinel (star, CF=39, WF=9, FP=81.2\%) sits above the break-even line, while LLM-as-Judge (diamond, CF=70, WF=94, FP=42.7\%) falls below it. Marker color encodes Flip Precision.}
\label{fig:tradeoff}
\end{subfigure}
\caption{Threshold sweep analysis: (a) Net Gain as a function of the decision threshold $\tau$; (b) Correct Flips vs.\ Wrong Flips along the same sweep.}
\Description{Panel (a): A line plot of Net Gain versus threshold tau from 0.1 to 0.9. Net Gain is strongly negative at low tau (reaching about minus 20 percent near tau 0.1), rises monotonically, crosses zero near tau 0.61, peaks at tau 0.81 with NG above +1 percent, and remains positive up to tau 0.95. A horizontal dashed line marks the per-dataset benchmark at +1.71 percent. Panel (b): A scatter plot of Correct Flips (vertical) versus Wrong Flips (horizontal). A dashed diagonal marks the break-even line where CF equals WF. A trajectory of circles traces the threshold sweep, color-coded by Flip Precision from red (low) to green (high). Sentinel is marked with a red star at CF=39 WF=9, clearly above the break-even line. LLM-as-Judge is marked with a blue diamond at CF=70 WF=94, below the break-even line.}
\label{fig:threshold_analysis}
\end{figure*}

The sweep reveals that NG\,$>$\,0 spans $\tau \in [0.61, 0.95]$, a width of 0.34, meaning that even without fine-grained tuning, Sentinel shows positive gains within this broad interval. The globally optimal fixed threshold is $\tau = 0.81$ yielding NG~=~+1.08\%. Compared to the per-dataset result of +1.71\%, the 0.63-percentage-point gap reflects the benefit of adaptive thresholds across datasets with varying Minority Truth rates. Figure~\ref{fig:threshold_analysis} visualizes the full sweep: (a)~plots Net Gain against $\tau$; (b)~shows the CF/WF trade-off, where Sentinel (CF=39, WF=9) sits above the break-even line with high precision and moderate recall, while LLM-as-Judge (CF=70, WF=94) falls below it, illustrating why Flip Precision, not Recovery Rate, determines intervention value.

\subsection{Classifier Comparison}

To verify LightGBM's optimality, we compare six classifiers under the same feature set and evaluation framework (Table~\ref{tab:classifier}).

LightGBM achieves the best NG (+1.71\%), with XGBoost (+1.60\%) and CatBoost (+1.54\%) following closely, collectively validating gradient boosting trees' advantage for small samples with heterogeneous features. Random Forest achieves the highest AUC (0.751) but only NG~=~+1.37\% due to less precise probability calibration. Logistic Regression (NG~=~+0.68\%) confirms significant nonlinear feature--label relationships. MLP performs worst (NG~=~+0.11\%, AUC~=~0.508), consistent with neural network overfitting on small samples.

\begin{table}[!htbp]
\centering
\caption{Classifier comparison under the same feature set and evaluation framework. The GBDT family outperforms all others.}
\label{tab:classifier}
\small
\setlength{\tabcolsep}{4pt}
\begin{tabular}{@{}clrrrrr@{}}
\toprule
\textbf{\#} & \textbf{Classifier} & \textbf{AUC} & \textbf{CF} & \textbf{WF} & \textbf{FP\%} & \textbf{NG\%} \\
\midrule
1 & LightGBM (Ours) & .741 & 39 & 9  & 81.2 & +1.71 \\
2 & XGBoost         & .743 & 37 & 9  & 80.4 & +1.60 \\
3 & CatBoost        & .726 & 37 & 10 & 78.7 & +1.54 \\
4 & Random Forest   & .751 & 33 & 9  & 78.6 & +1.37 \\
5 & Logistic Reg.   & .734 & 25 & 13 & 65.8 & +0.68 \\
6 & MLP             & .508 & 7  & 5  & 58.3 & +0.11 \\
\bottomrule
\end{tabular}
\end{table}

\subsection{Multi-Seed Stability}

LightGBM's training involves randomness (feature subsampling, data subsampling, split point selection). To assess stability, we run the complete pipeline with 20 random seeds (seed~=~0,\,1,\,\ldots,\,19).

\begin{figure}[!htbp]
\centering
\includegraphics[width=0.95\columnwidth]{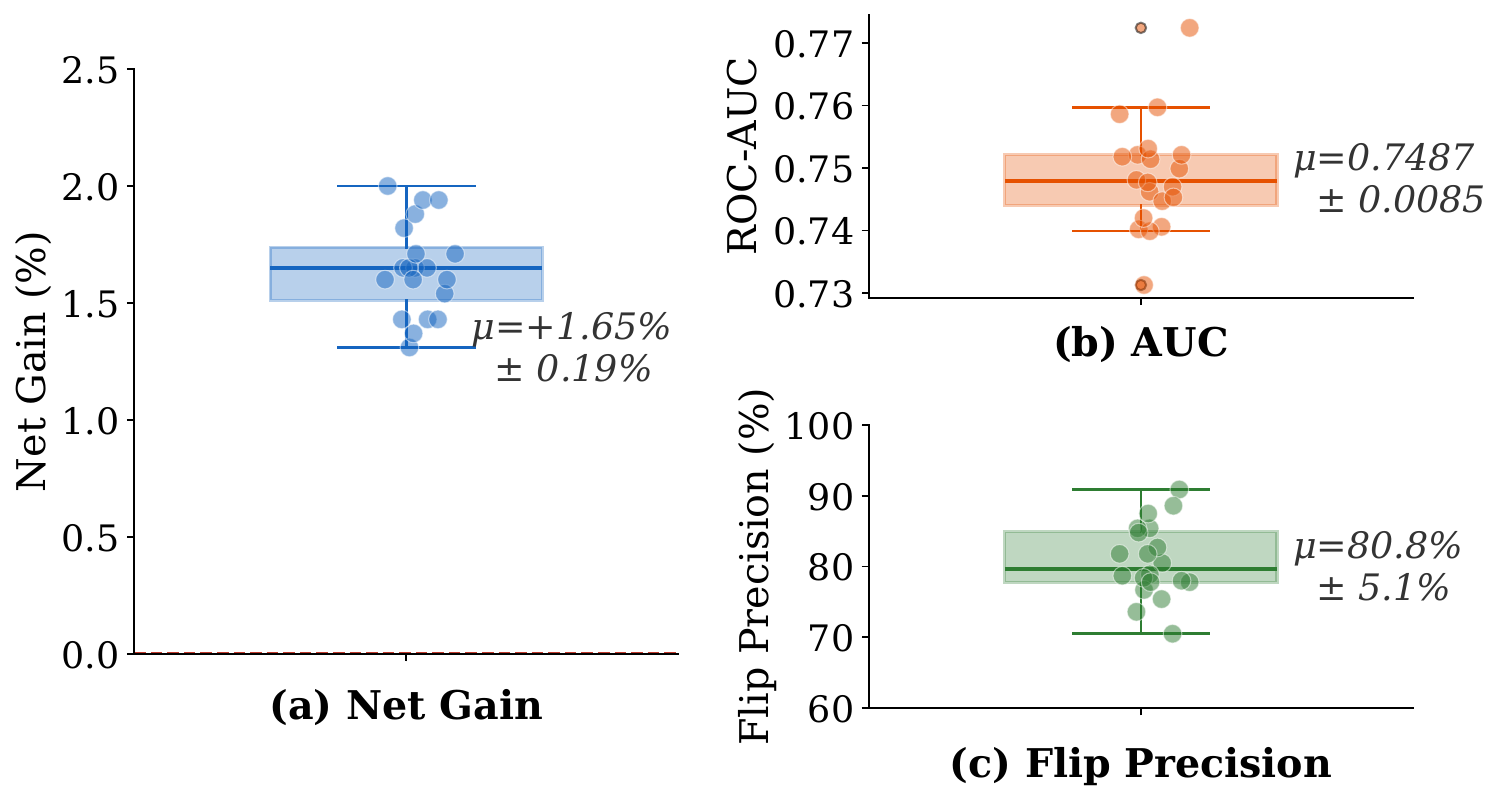}
\caption{20-seed stability. All seeds produce positive Net Gain (mean +1.65\%\,$\pm$\,0.19\%), confirming robustness to random initialization.}
\Description{Three box plots side by side for 20 random seeds. Left: Net Gain with mean 1.65 percent and all points above zero. Center: ROC-AUC with mean 0.7487. Right: Flip Precision with mean 80.8 percent. Individual data points are overlaid on each box plot.}
\label{fig:stability}
\end{figure}

The NG mean is +1.65\% (std~=~0.19\%), with minimum +1.31\% and maximum +2.00\%. All 20 seeds yield positive NG, demonstrating that Sentinel's positive effect is independent of random initialization. AUC mean is 0.749 (std~=~0.009), FP mean is 80.8\% (std~=~5.1\%). The seed~=~42 used in main results (NG~=~+1.71\%) falls within normal fluctuation range. Figure~\ref{fig:stability} presents the 20-seed box plots.

\subsection{Error Analysis}

Sentinel produces 9 wrong flips (WF) in the main results. We conduct qualitative analysis and categorize them into two error patterns.

\textbf{Pattern~A: Question Ambiguity (3 cases, 33.3\%).} These questions admit multiple reasonable interpretations, where majority and minority each provide different but individually reasonable answers. Sentinel tends to flip because the minority's argumentation presents stronger logical signals, but the ground truth annotation selects the alternative interpretation. These errors reflect benchmark annotation limitations rather than model judgment deficiencies.

\textbf{Pattern~B: True Hard Errors (6 cases, 66.7\%).} The majority holds the correct answer with reasonable argumentation, yet Sentinel still flips. The minority happens to exhibit behavioral patterns highly similar to genuine Minority Truth (consistently maintaining position, introducing seemingly independent arguments, or displaying high reasoning confidence), triggering misjudgment.

A noteworthy observation: Agent~A (GPT-4o-mini, ``Rigorous Auditor'') appears as the minority in 6 of 9 WFs, and 7 of 9 WFs come from MMLU-STEM, suggesting that in knowledge-intensive scenarios, erroneous minority arguments may ``masquerade'' as high-quality reasoning more easily.
\section{Discussion}
\label{sec:discussion}

\subsection{From Counting Heads to Auditing Evidence}

The core finding can be distilled into one proposition: at the aggregation layer of multi-agent debate, a paradigm shift from ``counting heads'' to ``auditing evidence'' is feasible through extremely lightweight means. \MS{} uses a multi-dimensional \emph{debate fingerprint} and a single LightGBM classifier, modifies no LLM weights, and achieves positive Net Gain on all six datasets with all 20 random seeds positive. The significance lies not in the absolute gain figure itself, but in demonstrating that debate logs contain behavioral signals sufficient to distinguish ``reliable consensus'' from ``spurious consensus.''

The failure of LLM-as-Judge (NG~=~$-$1.37\%) reinforces this from the opposite direction. \MS{} succeeds precisely because it steps outside the LLM's cognitive space, making meta-level judgments through statistical behavioral patterns rather than semantic content. This ``cognitive orthogonality'' may be a principle worth deeper exploration in future multi-agent system design.

A natural concern is that eight Semantic Audit features are themselves extracted by GPT-4o, seemingly reintroducing the very LLM dependency the framework aims to circumvent. We acknowledge this tension directly: if LLM shared blind spots are the core problem, then GPT-4o's quality assessments could carry systematic biases correlated with the debating agents' errors.

However, three observations mitigate this concern. First, the audit LLM serves as a \emph{structured feature extractor}, producing coarse-grained signals (binary flags and 1--5 ordinal scores) rather than directly adjudicating which answer is correct. Second, and most critically, ablation results (Table~\ref{tab:ablation}) show that removing all eight audit features reduces NG by only 0.29 percentage points, while Debate Dynamics alone recovers 73\% of the full model's net gain. The system's backbone is thus non-LLM behavioral statistics; the audit features provide a useful but non-essential supplement. Third, unlike LLM-as-Judge which makes a direct semantic judgment on answer correctness, the audit features are filtered through a tree-based classifier that learns to weight them against behavioral signals, reducing the impact of any individual biased assessment. In summary, ``cognitive orthogonality'' is best understood as a \emph{spectrum} rather than a binary property: \MS{} is not entirely free of LLM influence, but its decision-making center of gravity lies firmly in non-LLM behavioral patterns, which explains why it avoids the failure mode of LLM-as-Judge.

\subsection{Limitations}

This study has the following limitations. First, the debate system is fixed at 3 agents and 2 rounds, producing only 2:1 divergences; whether the framework generalizes to more complex structures (e.g., 5 agents, 3:2 splits) remains to be verified. Second, Semantic Audit features depend on additional GPT-4o calls, introducing non-trivial inference costs. Third, per-dataset threshold optimization requires labeled divergent samples, leaving zero-shot threshold calibration an open problem. Fourth, 686 divergent samples are relatively limited, especially for some datasets (e.g., ARC-Challenge with only 33 divergent samples and 14 Minority Truth cases), where individual flip decisions can significantly affect NG. Per-dataset threshold optimization on such small subsets carries overfitting risk: the positive NG observed on these datasets may partly reflect fortuitous threshold selection rather than robust discriminative ability. Future work should report bootstrap confidence intervals or adopt cross-validated threshold selection to quantify this uncertainty.

\subsection{Future Directions}

We consider the following directions worthy of exploration. First, combining \MS{} with Process Reward Models~\citep{lightman2024verify} to replace GPT-4o-dependent semantic audit features with step-level reasoning quality assessment. Second, exploring meta-learning frameworks for domain-adaptive threshold calibration. Third, extending to 5--7 agents to study more complex divergence structures and coalition dynamics. Fourth, validating in real-world scenarios such as RAG systems with conflicting retrieval paths or code generation with multiple candidate solutions.




\section{Conclusion}

This paper addresses the suppression of correct minority opinions under majority voting in multi-agent debate. Across six benchmarks, we identify the \emph{Minority Truth} phenomenon: 39.1\% of samples exhibit 2:1 splits, and in 25.5\% of these cases the minority is correct, yielding a theoretical recovery margin of 10.0 percentage points. To address this, we propose \textbf{\MS{}}, a plug-and-play \emph{Diagnosis--Cure} framework operating at the aggregation layer, which constructs a multi-dimensional \emph{debate fingerprint} (10-dim dynamics, 4-dim voting metadata, and 8-dim semantic audit) and trains a LightGBM meta-classifier to decide when to overturn majority voting. \MS{} achieves a Flip Precision of 81.2\% and consistent positive Net Gain across all six datasets (overall +1.71\%, CF~=~39, WF~=~9) and 20 random seeds (mean +1.65\%\,$\pm$\,0.19\%), while an LLM-as-Judge baseline degrades performance ($-$1.37\%), highlighting the advantage of behavioral-feature-based meta-classification. Overall, our results show that debate logs encode rich behavioral signals, motivating a shift from ``counting heads'' to ``auditing evidence,'' and positioning \MS{} as a lightweight and practical ``safety valve'' for recovering Minority Truth in multi-agent systems.

\bibliographystyle{ACM-Reference-Format}
\bibliography{references}


\begin{thebibliography}{22}


\ifx \showCODEN    \undefined \def \showCODEN     #1{\unskip}     \fi
\ifx \showISBNx    \undefined \def \showISBNx     #1{\unskip}     \fi
\ifx \showISBNxiii \undefined \def \showISBNxiii  #1{\unskip}     \fi
\ifx \showISSN     \undefined \def \showISSN      #1{\unskip}     \fi
\ifx \showLCCN     \undefined \def \showLCCN      #1{\unskip}     \fi
\ifx \shownote     \undefined \def \shownote      #1{#1}          \fi
\ifx \showarticletitle \undefined \def \showarticletitle #1{#1}   \fi
\ifx \showURL      \undefined \def \showURL       {\relax}        \fi
\providecommand\bibfield[2]{#2}
\providecommand\bibinfo[2]{#2}
\providecommand\natexlab[1]{#1}
\providecommand\showeprint[2][]{arXiv:#2}

\bibitem[Ai et~al\mbox{.}(2025)]%
        {ai2025beyond}
\bibfield{author}{\bibinfo{person}{Rui Ai}, \bibinfo{person}{Yuqi Pan},
  \bibinfo{person}{David Simchi-Levi}, \bibinfo{person}{Milind Tambe}, {and}
  \bibinfo{person}{Haifeng Xu}.} \bibinfo{year}{2025}\natexlab{}.
\newblock \showarticletitle{Beyond Majority Voting: {LLM} Aggregation by
  Leveraging Higher-Order Information}.
\newblock \bibinfo{journal}{\emph{arXiv preprint arXiv:2510.01499}}
  (\bibinfo{year}{2025}).
\newblock


\bibitem[Chan et~al\mbox{.}(2024)]%
        {chan2024chateval}
\bibfield{author}{\bibinfo{person}{Chi-Min Chan}, \bibinfo{person}{Weize Chen},
  \bibinfo{person}{Yusheng Su}, \bibinfo{person}{Jianxuan Yu},
  \bibinfo{person}{Wei Xue}, \bibinfo{person}{Shanghang Zhang},
  \bibinfo{person}{Jie Fu}, {and} \bibinfo{person}{Zhiyuan Liu}.}
  \bibinfo{year}{2024}\natexlab{}.
\newblock \showarticletitle{{ChatEval}: Towards Better {LLM}-based Evaluators
  through Multi-Agent Debate}. In \bibinfo{booktitle}{\emph{Proceedings of the
  12th International Conference on Learning Representations (ICLR)}}.
\newblock


\bibitem[Chang et~al\mbox{.}(2025)]%
        {chang2025main}
\bibfield{author}{\bibinfo{person}{Chia-Yuan Chang}, \bibinfo{person}{Zhimeng
  Jiang}, \bibinfo{person}{Vineeth Rakesh}, \bibinfo{person}{Menghai Pan},
  \bibinfo{person}{Chin-Chia~Michael Yeh}, \bibinfo{person}{Guanchu Wang},
  \bibinfo{person}{Mingzhi Hu}, \bibinfo{person}{Zhichao Xu},
  \bibinfo{person}{Yan Zheng}, \bibinfo{person}{Mahashweta Das}, {and}
  \bibinfo{person}{Na Zou}.} \bibinfo{year}{2025}\natexlab{}.
\newblock \showarticletitle{{MAIN-RAG}: Multi-Agent Filtering
  Retrieval-Augmented Generation}. In \bibinfo{booktitle}{\emph{Proceedings of
  the 63rd Annual Meeting of the Association for Computational Linguistics
  (ACL)}}. \bibinfo{pages}{2607--2622}.
\newblock


\bibitem[Choi et~al\mbox{.}(2025)]%
        {choi2025debate}
\bibfield{author}{\bibinfo{person}{Hyeong~Kyu Choi}, \bibinfo{person}{Xiaojin
  Zhu}, {and} \bibinfo{person}{Sharon Li}.} \bibinfo{year}{2025}\natexlab{}.
\newblock \showarticletitle{Debate or Vote: Which Yields Better Decisions in
  Multi-Agent Large Language Models?}. In \bibinfo{booktitle}{\emph{Advances in
  Neural Information Processing Systems (NeurIPS)}}.
\newblock


\bibitem[Clark et~al\mbox{.}(2018)]%
        {clark2018arc}
\bibfield{author}{\bibinfo{person}{Peter Clark}, \bibinfo{person}{Isaac
  Cowhey}, \bibinfo{person}{Oren Etzioni}, \bibinfo{person}{Tushar Khot},
  \bibinfo{person}{Ashish Sabharwal}, \bibinfo{person}{Carissa Schoenick},
  {and} \bibinfo{person}{Oyvind Tafjord}.} \bibinfo{year}{2018}\natexlab{}.
\newblock \showarticletitle{Think You Have Solved Question Answering? {Try
  ARC}, the {AI2} Reasoning Challenge}.
\newblock \bibinfo{journal}{\emph{arXiv preprint arXiv:1803.05457}}
  (\bibinfo{year}{2018}).
\newblock


\bibitem[Cobbe et~al\mbox{.}(2021)]%
        {cobbe2021gsm8k}
\bibfield{author}{\bibinfo{person}{Karl Cobbe}, \bibinfo{person}{Vineet
  Kosaraju}, \bibinfo{person}{Mohammad Bavarian}, \bibinfo{person}{Mark Chen},
  \bibinfo{person}{Heewoo Jun}, \bibinfo{person}{Lukasz Kaiser},
  \bibinfo{person}{Matthias Plappert}, \bibinfo{person}{Jerry Tworek},
  \bibinfo{person}{Jacob Hilton}, \bibinfo{person}{Reiichiro Nakano},
  \bibinfo{person}{Christopher Hesse}, {and} \bibinfo{person}{John Schulman}.}
  \bibinfo{year}{2021}\natexlab{}.
\newblock \showarticletitle{Training Verifiers to Solve Math Word Problems}.
\newblock \bibinfo{journal}{\emph{arXiv preprint arXiv:2110.14168}}
  (\bibinfo{year}{2021}).
\newblock


\bibitem[Du et~al\mbox{.}(2024)]%
        {du2024improving}
\bibfield{author}{\bibinfo{person}{Yilun Du}, \bibinfo{person}{Shuang Li},
  \bibinfo{person}{Antonio Torralba}, \bibinfo{person}{Joshua~B. Tenenbaum},
  {and} \bibinfo{person}{Igor Mordatch}.} \bibinfo{year}{2024}\natexlab{}.
\newblock \showarticletitle{Improving Factuality and Reasoning in Language
  Models through Multiagent Debate}. In \bibinfo{booktitle}{\emph{Proceedings
  of the 41st International Conference on Machine Learning (ICML)}}.
\newblock


\bibitem[Estornell and Liu(2024)]%
        {estornell2024multi}
\bibfield{author}{\bibinfo{person}{Andrew Estornell} {and}
  \bibinfo{person}{Yang Liu}.} \bibinfo{year}{2024}\natexlab{}.
\newblock \showarticletitle{Multi-{LLM} Debate: Framework, Principles, and
  Interventions}. In \bibinfo{booktitle}{\emph{Advances in Neural Information
  Processing Systems (NeurIPS)}}.
\newblock


\bibitem[Hendrycks et~al\mbox{.}(2021)]%
        {hendrycks2021mmlu}
\bibfield{author}{\bibinfo{person}{Dan Hendrycks}, \bibinfo{person}{Collin
  Burns}, \bibinfo{person}{Steven Basart}, \bibinfo{person}{Andy Zou},
  \bibinfo{person}{Mantas Mazeika}, \bibinfo{person}{Dawn Song}, {and}
  \bibinfo{person}{Jacob Steinhardt}.} \bibinfo{year}{2021}\natexlab{}.
\newblock \showarticletitle{Measuring Massive Multitask Language
  Understanding}. In \bibinfo{booktitle}{\emph{Proceedings of the 9th
  International Conference on Learning Representations (ICLR)}}.
\newblock


\bibitem[Hu et~al\mbox{.}(2025)]%
        {hu2025multi}
\bibfield{author}{\bibinfo{person}{Tianyu Hu}, \bibinfo{person}{Zixiang Tan},
  \bibinfo{person}{Shuaiqi Wang}, \bibinfo{person}{Huiying Qu}, {and}
  \bibinfo{person}{Tianyi Chen}.} \bibinfo{year}{2025}\natexlab{}.
\newblock \showarticletitle{Multi-Agent Debate for {LLM} Judges with Adaptive
  Stability Detection}. In \bibinfo{booktitle}{\emph{Advances in Neural
  Information Processing Systems (NeurIPS)}}.
\newblock


\bibitem[Ke et~al\mbox{.}(2017)]%
        {ke2017lightgbm}
\bibfield{author}{\bibinfo{person}{Guolin Ke}, \bibinfo{person}{Qi Meng},
  \bibinfo{person}{Thomas Finley}, \bibinfo{person}{Taifeng Wang},
  \bibinfo{person}{Wei Chen}, \bibinfo{person}{Weidong Ma},
  \bibinfo{person}{Qiwei Ye}, {and} \bibinfo{person}{Tie-Yan Liu}.}
  \bibinfo{year}{2017}\natexlab{}.
\newblock \showarticletitle{{LightGBM}: A Highly Efficient Gradient Boosting
  Decision Tree}. In \bibinfo{booktitle}{\emph{Advances in Neural Information
  Processing Systems (NeurIPS)}}, Vol.~\bibinfo{volume}{30}.
  \bibinfo{pages}{3146--3154}.
\newblock


\bibitem[Kim et~al\mbox{.}(2025)]%
        {kim2025correlated}
\bibfield{author}{\bibinfo{person}{Elliot Kim}, \bibinfo{person}{Avi Garg},
  \bibinfo{person}{Kenny Peng}, {and} \bibinfo{person}{Nikhil Garg}.}
  \bibinfo{year}{2025}\natexlab{}.
\newblock \showarticletitle{Correlated Errors in Large Language Models}. In
  \bibinfo{booktitle}{\emph{Proceedings of the 42nd International Conference on
  Machine Learning (ICML)}}.
\newblock


\bibitem[Liang et~al\mbox{.}(2024)]%
        {liang2024encouraging}
\bibfield{author}{\bibinfo{person}{Tian Liang}, \bibinfo{person}{Zhiwei He},
  \bibinfo{person}{Wenxiang Jiao}, \bibinfo{person}{Xing Wang},
  \bibinfo{person}{Yan Wang}, \bibinfo{person}{Rui Wang},
  \bibinfo{person}{Yujiu Yang}, \bibinfo{person}{Shuming Shi}, {and}
  \bibinfo{person}{Zhaopeng Tu}.} \bibinfo{year}{2024}\natexlab{}.
\newblock \showarticletitle{Encouraging Divergent Thinking in Large Language
  Models through Multi-Agent Debate}. In \bibinfo{booktitle}{\emph{Proceedings
  of the 2024 Conference on Empirical Methods in Natural Language Processing
  (EMNLP)}}. \bibinfo{pages}{17889--17904}.
\newblock


\bibitem[Lightman et~al\mbox{.}(2024)]%
        {lightman2024verify}
\bibfield{author}{\bibinfo{person}{Hunter Lightman}, \bibinfo{person}{Vineet
  Kosaraju}, \bibinfo{person}{Yura Burda}, \bibinfo{person}{Harri Edwards},
  \bibinfo{person}{Bowen Baker}, \bibinfo{person}{Teddy Lee},
  \bibinfo{person}{Jan Leike}, \bibinfo{person}{John Schulman},
  \bibinfo{person}{Ilya Sutskever}, {and} \bibinfo{person}{Karl Cobbe}.}
  \bibinfo{year}{2024}\natexlab{}.
\newblock \showarticletitle{Let's Verify Step by Step}. In
  \bibinfo{booktitle}{\emph{Proceedings of the 12th International Conference on
  Learning Representations (ICLR)}}.
\newblock


\bibitem[Lin et~al\mbox{.}(2022)]%
        {lin2022truthfulqa}
\bibfield{author}{\bibinfo{person}{Stephanie Lin}, \bibinfo{person}{Jacob
  Hilton}, {and} \bibinfo{person}{Owain Evans}.}
  \bibinfo{year}{2022}\natexlab{}.
\newblock \showarticletitle{{TruthfulQA}: Measuring How Models Mimic Human
  Falsehoods}. In \bibinfo{booktitle}{\emph{Proceedings of the 60th Annual
  Meeting of the Association for Computational Linguistics (ACL)}}.
  \bibinfo{pages}{3214--3252}.
\newblock


\bibitem[Moscovici(1976)]%
        {moscovici1976social}
\bibfield{author}{\bibinfo{person}{Serge Moscovici}.}
  \bibinfo{year}{1976}\natexlab{}.
\newblock \bibinfo{booktitle}{\emph{Social Influence and Social Change}}.
\newblock \bibinfo{publisher}{Academic Press}, \bibinfo{address}{London}.
\newblock


\bibitem[Sakaguchi et~al\mbox{.}(2021)]%
        {sakaguchi2021winogrande}
\bibfield{author}{\bibinfo{person}{Keisuke Sakaguchi},
  \bibinfo{person}{Ronan~Le Bras}, \bibinfo{person}{Chandra Bhagavatula}, {and}
  \bibinfo{person}{Yejin Choi}.} \bibinfo{year}{2021}\natexlab{}.
\newblock \showarticletitle{{WinoGrande}: An Adversarial {Winograd} Schema
  Challenge at Scale}.
\newblock \bibinfo{journal}{\emph{Commun. ACM}} \bibinfo{volume}{64},
  \bibinfo{number}{9} (\bibinfo{year}{2021}), \bibinfo{pages}{99--106}.
\newblock


\bibitem[Talmor et~al\mbox{.}(2019)]%
        {talmor2019commonsenseqa}
\bibfield{author}{\bibinfo{person}{Alon Talmor}, \bibinfo{person}{Jonathan
  Herzig}, \bibinfo{person}{Nicholas Lourie}, {and} \bibinfo{person}{Jonathan
  Berant}.} \bibinfo{year}{2019}\natexlab{}.
\newblock \showarticletitle{{CommonsenseQA}: A Question Answering Challenge
  Targeting World Knowledge}. In \bibinfo{booktitle}{\emph{Proceedings of the
  2019 Conference of the North American Chapter of the Association for
  Computational Linguistics (NAACL)}}.
\newblock


\bibitem[Wang et~al\mbox{.}(2023)]%
        {wang2023selfconsistency}
\bibfield{author}{\bibinfo{person}{Xuezhi Wang}, \bibinfo{person}{Jason Wei},
  \bibinfo{person}{Dale Schuurmans}, \bibinfo{person}{Quoc Le},
  \bibinfo{person}{Ed Chi}, \bibinfo{person}{Sharan Narang},
  \bibinfo{person}{Aakanksha Chowdhery}, {and} \bibinfo{person}{Denny Zhou}.}
  \bibinfo{year}{2023}\natexlab{}.
\newblock \showarticletitle{Self-Consistency Improves Chain of Thought
  Reasoning in Language Models}. In \bibinfo{booktitle}{\emph{Proceedings of
  the 11th International Conference on Learning Representations (ICLR)}}.
\newblock


\bibitem[Wu et~al\mbox{.}(2025)]%
        {wu2025debate}
\bibfield{author}{\bibinfo{person}{Haolun Wu}, \bibinfo{person}{Zhenkun Li},
  {and} \bibinfo{person}{Lingyao Li}.} \bibinfo{year}{2025}\natexlab{}.
\newblock \showarticletitle{Can {LLM} Agents Really Debate? {A} Controlled
  Study of Multi-Agent Debate in Logical Reasoning}.
\newblock \bibinfo{journal}{\emph{arXiv preprint arXiv:2511.07784}}
  (\bibinfo{year}{2025}).
\newblock


\bibitem[Yang et~al\mbox{.}(2026)]%
        {yang2026agentauditor}
\bibfield{author}{\bibinfo{person}{Wei Yang}, \bibinfo{person}{Shixuan Li},
  \bibinfo{person}{Heng Ping}, \bibinfo{person}{Peiyu Zhang},
  \bibinfo{person}{Paul Bogdan}, {and} \bibinfo{person}{Jesse Thomason}.}
  \bibinfo{year}{2026}\natexlab{}.
\newblock \showarticletitle{Auditing Multi-Agent {LLM} Reasoning Trees
  Outperforms Majority Vote and {LLM}-as-Judge}.
\newblock \bibinfo{journal}{\emph{arXiv preprint arXiv:2602.09341}}
  (\bibinfo{year}{2026}).
\newblock


\bibitem[Zheng et~al\mbox{.}(2023)]%
        {zheng2023judging}
\bibfield{author}{\bibinfo{person}{Lianmin Zheng}, \bibinfo{person}{Wei-Lin
  Chiang}, \bibinfo{person}{Ying Sheng}, \bibinfo{person}{Siyuan Zhuang},
  \bibinfo{person}{Zhanghao Wu}, \bibinfo{person}{Yonghao Zhuang},
  \bibinfo{person}{Zi Lin}, \bibinfo{person}{Zhuohan Li},
  \bibinfo{person}{Dacheng Li}, \bibinfo{person}{Eric Xing},
  \bibinfo{person}{Hao Zhang}, \bibinfo{person}{Joseph~E. Gonzalez}, {and}
  \bibinfo{person}{Ion Stoica}.} \bibinfo{year}{2023}\natexlab{}.
\newblock \showarticletitle{Judging {LLM}-as-a-Judge with {MT}-Bench and
  Chatbot Arena}. In \bibinfo{booktitle}{\emph{Advances in Neural Information
  Processing Systems (NeurIPS)}}, Vol.~\bibinfo{volume}{36}.
  \bibinfo{pages}{46595--46623}.
\newblock


\end{thebibliography}

\appendix
\section{Prompt Templates}
\label{app:prompts}

{\small
This appendix provides the prompt templates used in our experiments. Agent and debate prompts are presented verbatim; the Semantic Audit and LLM-as-Judge prompts are summarized for brevity.

\subsection{Agent System Prompts}
All three agents use temperature~=~0.7; cognitive diversity arises from different vendor models and the role assignments below.

\smallskip\noindent\textbf{Agent~A --- Rigorous Auditor} (GPT-4o-mini): \emph{``You are a meticulous fact-checker and logical auditor. You carefully verify each step of reasoning before accepting any conclusion. You are skeptical of groupthink and do NOT change your answer unless you find a genuine logical error in your own reasoning.''}

\smallskip\noindent\textbf{Agent~B --- Balanced Analyst} (Gemini-2.0-Flash): \emph{``You are a balanced analytical thinker. You consider all perspectives fairly and weigh evidence carefully. You are open to changing your answer when presented with a genuinely stronger argument, but you require substantive reasoning --- not mere popularity --- to change your mind.''}

\smallskip\noindent\textbf{Agent~C --- Intuitive Challenger} (Claude Haiku~4.5): \emph{``You are an intuitive problem solver who thinks outside the box. You often spot unconventional solutions that others miss. However, you can sometimes be swayed by confident-sounding arguments from others.''}

\subsection{Debate Prompts}

\noindent\textit{Round~0 (Independent).}
\begin{quote}\footnotesize\raggedright\ttfamily
You are solving the following problem. Think step by step and show your reasoning clearly. Problem: \{question\}\par
Provide your answer in the following format:\par
\textnormal{$\langle$}\allowbreak reasoning\textnormal{$\rangle$}...\textnormal{$\langle$/}reasoning\textnormal{$\rangle$} \textnormal{$\langle$}answer\textnormal{$\rangle$}...\textnormal{$\langle$/}answer\textnormal{$\rangle$}
\end{quote}

\noindent\textit{Rounds~1--2 (Debate).}
\begin{quote}\footnotesize\raggedright\ttfamily
Problem: \{question\}\par
Your previous answer was: \{own\_previous\}\par
Other participants gave the following responses: \{others\}\par
Consider the other participants' reasoning carefully. If you find errors in their logic, point them out specifically. Do NOT change your answer just because others disagree --- only change if you find a concrete logical reason.\par
Respond in format: \textnormal{$\langle$}reasoning\textnormal{$\rangle$}...\textnormal{$\langle$/}reasoning\textnormal{$\rangle$} \textnormal{$\langle$}answer\textnormal{$\rangle$}...\textnormal{$\langle$/}answer\textnormal{$\rangle$} \textnormal{$\langle$}stance\_change\textnormal{$\rangle$}YES or NO\textnormal{$\langle$/}stance\_change\textnormal{$\rangle$} \textnormal{$\langle$}change\_reason\textnormal{$\rangle$}...\textnormal{$\langle$/}change\_reason\textnormal{$\rangle$}
\end{quote}

\subsection{Semantic Audit Prompt}
GPT-4o (temperature~=~0.0) acts as an external logic auditor, evaluating reasoning quality of both sides without knowledge of the correct answer. The system prompt instructs it to output only a valid JSON object. The user prompt provides the problem, majority reasoning texts, and minority reasoning texts, then requests seven structured fields:
\begin{quote}\footnotesize\raggedright
\feat{minority\_new\_evidence}, \feat{majority\_echo\_chamber}, \feat{minority\_finds\_error}, \feat{majority\_logical\_gap},\\
\feat{minority\_reasoning\_score}~(1--5), \feat{majority\_reasoning\_score}~(1--5), and \feat{blind\_follower\_count}~(0--2).
\end{quote}
The eighth feature, \feat{reasoning\_score\_diff}, is computed client-side.

\subsection{LLM-as-Judge Baseline Prompt}
For the LLM-as-Judge baseline, GPT-4o (temperature~=~0.0) reads the complete debate transcript and is instructed to determine the correct answer based on reasoning quality rather than vote count, outputting only the answer value.
}

\end{document}